\documentclass[12pt]{article}

\usepackage{aaspp4,natbib}

\def\la{\mathrel{\hbox{\rlap{\hbox{\lower4pt\hbox{$\sim$}}}\hbox{$<$}}}}
\def\ga{\mathrel{\hbox{\rlap{\hbox{\lower4pt\hbox{$\sim$}}}\hbox{$>$}}}}

\newcommand{\be}{\begin{eqnarray}}
\newcommand{\ee}{\end{eqnarray}}

\newcommand{\msol}{\ifmmode{{\rm M}_\odot}\else{M$_\odot$}\fi}
\newcommand{\foe}{\ifmmode{10^{51}}\else{$10^{51}$}\fi}

\newcommand{\xni}{\ifmmode{{\rm X}_{\rm Ni}}\else{X$_{\rm Ni}$}\fi}
\def\ang{\hbox{\AA}}
\def\Teff{\ifmmode{T_{\rm eff}}\else{\hbox{$T_{\rm eff}$} }\fi}
\def\Rzero{\ifmmode{R_0}\else{\hbox{$R_0$} }\fi}

\def\SP2{{\tt IBM SP2}}

\def\PC2{{\tt PC$^2$}}

\def\logg{\log(g)}
\def\mh{[{\rm M/H}]}

\def\inu{\ifmmode{I_{\nu}}\else{\hbox{$I_{\nu}$} }\fi}
\def\snu{\ifmmode{S_{\nu}}\else{\hbox{$S_{\nu}$} }\fi}
\def\jnu{\ifmmode{J_{\nu}}\else{\hbox{$J_{\nu}$} }\fi}

\def\fep{\ifmmode{{\rm Fe II}}\else\hbox{Fe~II }\fi}

  at 12.0 true pt %scaled \magstep1

\def\phoenix{{\tt PHOENIX}}

\def\phoenix{{\tt PHOENIX}}

\def\b{\beta}

\def\l{\lambda}
\def\L{\Lambda}
\def\t{\tau}

\def\rout{\ifmmode{r_{\rm out}}\else\hbox{$r_{\rm out}$}\fi}
\def\tmax{\ifmmode{\tau_{\rm max}}\else\hbox{$\tau_{\rm max}$}\fi}
\def\tstd{\ifmmode{\tau_{\rm std}}\else\hbox{$\tau_{\rm std}$}\fi}
\def\vmax{\ifmmode{v_{\rm max}}\else\hbox{$v_{\rm max}$}\fi}
\def\muE{\ifmmode{\mu_{\rm E}}\else\hbox{$\mu_{\rm E}$}\fi} 
\def\pE{\ifmmode{p_{\rm E}}\else\hbox{$p_{\rm E}$}\fi} 
\def\bmax{\ifmmode{\b_{\rm max}}\else\hbox{$\b_{\rm max}$}\fi}

\def\ang{\hbox{\AA}}

\def\Msun{\hbox{$\,$M$_\odot$} }

\def\Teff{\hbox{$\,T_{\rm eff}$} }

\def\rout{\hbox{$r_{\rm out}$} }

\def\chistd{\ifmmode{\chi_{\rm std}}\else\hbox{$\chi_{\rm std}$}\fi}

\def\K{\,{\rm K}}
\def\msol{$M_\odot$}
\def\foe{10^{51}}

\def\xni{{\rm X}_{\rm Ni}}

\def\lstar{\ifmmode{\Lambda^*}\else\hbox{$\Lambda^*$}\fi} 
\def\Rop{\ifmmode{[R_{ij}]}\else\hbox{$[R_{ij}]$}\fi}

\def\Rji{\ifmmode{[R_{ji}]}\else\hbox{$[R_{ji}]$}\fi}
\def\Rstar{\ifmmode{[R_{ij}^*]}\else\hbox{$[R_{ij}^*]$}\fi}

\def\Rjistar{\ifmmode{[R_{ji}^*]}\else\hbox{$[R_{ji}^*]$}\fi}
\def\DRji{\ifmmode{[\Delta R_{ji}]}\else\hbox{$[\Delta R_{ji}]$}\fi}
\def\DRij{\ifmmode{[\Delta R_{ij}]}\else\hbox{$[\Delta R_{ij}]$}\fi}

\def\ns{\ifmmode{N_{\rm s}}          % Anzahl der tau-punkte
        \else\hbox{$N_{\rm s}$}\fi}

\def\mat#1{{\bf #1}}     % Macro fr Matrizen
\def\vek#1{{#1}}         % Macro fr Vektoren

\newcount\eqcount
\def
  \nummer{
    \global\advance\eqcount by 1
    (\the\eqcount)
  }

\def
  \numadv{
    \global\advance\eqcount by 1
  }

\def
   \numout#1{
     (\the\eqcount #1)
  }

\def\ivek#1#2{\ifmmode{\vek{I}^{#1}_{#2}}
        \else\hbox{$\vek{I}^{#1}_{#2}$}\fi}

\def\tmat#1#2{\ifmmode{\mat{t}^{#1}_{#2}}
        \else\hbox{$\mat{t}^{#1}_{#2}$}\fi}
\def\rmat#1#2{\ifmmode{\mat{r}^{#1}_{#2}}
        \else\hbox{$\mat{r}^{#1}_{#2}$}\fi}
\def\bvek#1#2{\ifmmode{\beta^{#1}_{#2}}
        \else\hbox{$\beta^{#1}_{#2}$}\fi}

\def\lp{\ifmmode{\lambda^+_\tau}           % lambda +
        \else\hbox{$\lambda^+_\tau$}\fi}
\def\lm{\ifmmode\lambda^-_\tau             % lambda -
        \else\hbox{$\lambda^-_\tau$}\fi}

\chardef\tilt=126
\begin{document}
\bibliographystyle{apj}

\title{The NextGen Model Atmosphere grid:\\
II. Spherically symmetric model atmospheres for giant stars 
with effective temperatures between 3000 and 6800~K}

\author{Peter H. Hauschildt}
\affil{Dept.\ of Physics and Astronomy \& Center for Simulational Physics, 
University of Georgia, Athens, GA 30602-2451\\
Email: {\tt yeti@hal.physast.uga.edu}}
\author{France Allard}
\affil{C.R.A.L (UML 5574) Ecole Normale Superieure, 69364 Lyon Cedex 7, France\\
E-Mail: \tt fallard@ens-lyon.fr}
\author{Jason Ferguson}
\affil{Physics Dept.,
Wichita State University,
Wichita, KS 67260-0032\\
Email: \tt jfergus@twsu.edu
}
\author{E. Baron}
\affil{Dept. of Physics and Astronomy, University of Oklahoma,\\
440 W. Brooks, Rm 131, Norman, OK 73019-0225\\
E-Mail: \tt baron@mail.nhn.ou.edu}
\author{David R. Alexander}
\affil{Physics Dept.,
Wichita State University,
Wichita, KS 67260-0032\\
Email: \tt dra@twsuvm.uc.twsu.edu
}
\medskip
{\rm ApJ, in press (November 1999).\\
Also available at {\tt ftp://calvin.physast.uga.edu/pub/preprints/}
and as HTML at {\tt http://dilbert.physast.uga.edu/\tilt yeti/PAPERS}.}

\begin{abstract}

We present the extension of our NextGen model atmosphere 
grid to the regime of giant stars. The input physics of the
models presented here is nearly identical to the NextGen dwarf atmosphere
models, however spherical geometry is used self-consistently
in the model calculations (including the radiative transfer).
We re-visit the discussion of the effects of spherical geometry on
the structure of the atmospheres and the emitted spectra and 
discuss the results of NLTE calculations for a few selected models.

\end{abstract}

\section{Introduction}

In a recent paper \cite[][hereafter: NG]{ng-hot} we have presented a grid of
plane parallel model atmospheres for dwarf stars ($\logg \geq 3.5$) and in the
temperature range $3000\leq \Teff \leq 10000\K$. The NG models are intended for
the analysis and modeling of main sequence stars and, where applicable,
sub-giants. Spherical extension effects become more important for smaller
gravities, which is the reason why the NG-grid stopped at $\logg = 3.5$. In
this paper, we present an extension of the NG grid to smaller gravities by
including the effects of spherical geometry in the model calculations. This
includes, for example, the calculation of the hydrostatic structure in
spherical geometry and the effects of spherical radiative transfer on the model
atmospheres.

In order to retain compatibility with the NG grid, the NG-giant grid presented
here uses the same parameterization of the element abundances as the NG grid.
Therefore, the models presented here are not applicable to chemically peculiar
stars such as, e.g., Carbon stars. However, there are many possible
applications for these models. For example, the Period-luminosity (PL)
relation of Cepheids is of fundamental importance for the determination of the
extragalactic distance scale.  \cite{alibert99} have used the atmosphere models
presented in this paper to connect evolutionary models to observed colors for
Cepheids with very good results.

In the next section we give  a brief overview over the model construction 
and the differences from the NG grid. Then we discuss the results, in particular
the effects of spherical symmetry and we end with a summary of the paper.

\section{Model calculations}

We have calculated the models presented in this paper using our
multipurpose model atmosphere code \phoenix, version 10.5. Details of the code and the
general input physics setup are discussed in \cite{ng-hot} and \cite{jcam} and
references therein. The model atmospheres for Cepheids
presented here were calculated with the same general input physics. However,
they use spherical geometry (including spherically symmetric radiative
transfer) rather than plane parallel geometry.
For giant models with low gravities ($\logg \le 3.5$), this can be
important for the correct calculation of the structure of the model
atmosphere and the synthetic spectrum \cite[c.f., e.g.][for a discussion of
spherical effects in B giants]{epscma,betacma}. 

Our combined molecular line list includes about 500 million molecular lines.
These lines are treated with a direct opacity sampling technique where each
line has its individual Voigt (for strong lines) or Gauss (weak lines) line
profile. The lines are selected for every model from the master line list at
the beginning of each model iteration to account for changes in the model
structure \cite[see][or details of the line selection process]{ng-hot}.
This procedure selects about 190
million molecular lines for a typical giant model atmosphere with $\Teff\approx 3000\K$.
Accordingly, we generally use the parallelized version of \phoenix\
\cite[]{parapap,parapap2} to perform calculations efficiently on parallel
supercomputers.  Details of the calculations are given in the above references
and are not repeated here.

The parameterization of giant models requires an additional parameter compared
to plane parallel model atmospheres. This complicates the task of relating
theoretical models to observed data, see \cite{bsw91} for a discussion of these
issues. In the models presented here, we set the stellar radius $R$ by the
condition $g_{\rm grav}=GM/R^2$, where we define $g_{\rm grav}$ as the
gravitational acceleration at $\tstd=1$, $G$ is the constant of gravity, and
$\tstd$ is the optical depth in the continuum at $1.2\,\mu$.  The luminosity
$L$ of the model is then given by $L=4\pi R^2 \sigma \Teff^4$. For convenience,
our model grid is based on the set of parameters $(\Teff,\logg,M,\mh)$. The
above formulae and the structures of the model atmospheres can be used to
transform them to any target set of parameters (e.g., for a different
definition of $\tstd$).

\subsection{Temperature correction procedure}

We iterate for the temperature structure of the atmosphere using a
generalization of the Uns\"old-Lucy temperature correction scheme to spherical
geometry and NLTE model calculations.  This has proven to work very well even
in extreme NLTE cases such as nova and supernova atmospheres. The temperature
correction procedure also requires virtually no memory and CPU time overhead.
The Uns\"old-Lucy correction scheme \cite[see][for a discussion of this and
other temperature correction schemes]{mihalas70}, uses the global constraint
equation of energy conservation to find corrections to the temperature that
will fulfill energy conservation better than the previous estimate.  We
have found it to be more stable than a Newton-Raphson linearization scheme and
it allows us to separate the temperature corrections from the statistical
equations. The latter property is extremely useful in NLTE calculations
where it allows us to use a nested iteration scheme for the energy
conservation and the statistical equilibrium equations. This significantly 
increases numerical stability and allows us to use much larger model
atoms than more conventional methods.

To derive the Uns\"old-Lucy correction, one uses the fact that
the {\em ratios} of the wavelength averaged absorption and
extinction coefficients 
\begin{eqnarray}
\kappa_P & = & \left(\int_0^\infty \kappa_\l B_l\,d\l\right)/B \\
\kappa_J & = & \left(\int_0^\infty \kappa_\l J_l\,d\l\right)/J \\
\chi_J & = & \left(\int_0^\infty \chi_\l F_l\,d\l\right)/F \\
\end{eqnarray}
(where $B,J,F$ denote the wavelength integrated Planck function, mean intensity
and radiation flux, respectively) depend much less on values of the
independent variables than do the averages themselves. 

Dropping terms of order $(v/c)$, one can then
use the angular moments of the radiative transfer equation to show that in order to
obtain radiation equilibrium $B$ should be corrected by an amount
\be
\delta B(r) &=& \frac{1}{\kappa_P}\big\{\kappa_J J - \kappa_P B+ \dot S/(4
\pi)\big\} \label{uleqn}\\ &&-\frac{\kappa_J}{\kappa_P} \big\{ 2(H(\tau=0)-H_0(\tau=0))
\nonumber\\
&&-
\frac{1}{fqr^2}\int_r^R q\,r'^2\, \chi_F 
\,(H(r')-H_0(r'))\,dr'\big\},\nonumber
\ee
where $H\equiv F/4\pi$, $H_0(\tau)$ is the value of the target luminosity at
that particular depth point (variable due to the velocity terms in comoving frame
radiative transfer calculations
and non-mechanical energy sources, the total {\em observed} luminosity $H_0(0)$
is an input parameter), Here, $q$ is the ``sphericity factor'' given
by
 \[ q =
\frac{1}{r^2}\exp\left(\int_{r_{\rm core}}^r\frac{3f - 1}{r'f}\, dr'\right), \] 
where $r_{\rm core}$ is the inner radius of the atmosphere, $R$ is the
total radius, $f(\t)=K(\t)/J(\t)$ is the ``Eddington
factor'', and $K=\int 
\mu^2 I \, d\mu$ is the second angular moment of the mean intensity. $\dot S$
describes all additional sources of energy such as mechanical energy supplied
by winds or non-thermal ionization due to $\gamma$--ray deposition. 

The first term in Eq.~\ref{uleqn} corresponds simply to a $\L$ iteration term
and will thus provide temperature corrections that are smaller than
required in the inner parts of the atmosphere, but will be accurate in
the outer, optically thin parts.
The second term of Eq.~\ref{uleqn}, however, is the dominant term in
the inner parts of the atmosphere. It provides a very good approximation to the
temperature corrections $\Delta T$ deep inside the atmosphere. Following
\cite{unsoeld55}, we have found that it is sometimes better to modify this general
scheme by excluding the contributions of extremely strong lines in the
opacity averages used for the calculations of $\Delta T$ because they tend to
dominate the average opacity but do not contribute as much to the total error
in the energy conservation constraint. 

\section{Results}

We have calculated a grid of solar abundance \cite[table 5 of][]{jaschek95} model
atmospheres.
In addition, we have calculated grids with metallicities $\mh=-0.3$, $-0.5$ and
$-0.7$ to facilitate comparison to observed LMC and SMC giants.  The models
span a range of $2000\K\leq \Teff \leq 6800\K$ and $0.0\leq\logg\leq 3.5$.  The
models are not available for all combinations of these parameters, e.g., at
high $\Teff$ and low $\logg$ the total radiative acceleration is larger than
the local gravity and hydrostatic models become invalid. In order to minimize
CPU time, the models were calculated for a fixed mass of $M=5\Msun$. In the
following, we discuss the effects of different model parameters on the resulting
structures and spectra. The effects of the model parameters on the colors 
are discussed in the context of evolutionary models by \cite{alibert99}.

\subsection{Structure of the atmospheres}

If Fig.~\ref{T-struc-logg} we show the temperature structure for models 
with varying $\logg$ for 3 effective temperatures. For the models with 
$\Teff=5600\K$, and the smallest gravity, the atmosphere does not become
convective within the maximum optical depth of the model, $\tstd=100$. 
The apparent flattening of the temperature drop in the outer layers of the
atmosphere is a result of the automatic ``compression'' of the optical depth scale
toward the outer edge of the model atmosphere. This means that the 
radii of the different optical depth points are much closer to each other in 
the outer atmosphere compared to $\tstd \approx 1$. This is most pronounced
in the coolest models shown, whereas the ``scale compression'' is nearly non-existent
for the model with $\Teff=5600\K$. We have also compared model atmospheres
calculated with the same parameters ($\Teff$, $\logg$) but for different
stellar masses. For $\Teff=3600\K$, $\logg=0.0$ and solar abundances,
we find only small temperature differences for stellar masses between $2.5$ and
$7.5\Msun$. However, the {\em absolute} differences in the synthetic spectra are equivalent
to a significant change in $\logg$ (see below) whereas the {\em relative}
differences in the spectra are small.
The radial extension of the atmospheres, here defined simply as the 
ratio of the outer radius $R_{\rm out}$ and the inner radius $R_{\rm in}$
of the atmosphere, is typically less than 20\% (see Fig.~\ref{extension}).
We have ignored all models that become unstable due to large radiative
accelerations (low $\logg$ models with very low or high temperatures), such
atmospheres have to be described by stellar wind models (Aufdenberg et al., in
preparation) and typically have much larger radial extensions. 

\subsection{Sphericity effects}

It is important to assess the relevance of the effects of spherical geometry
and radiative transfer on the synthetic spectra.  For hot stars, \cite{epscma,betacma}
have shown that a combination of geometric effects and line
blanketing can resolve a long standing problem of discrepancies between
synthetic and observed spectra.  The effects of spherical symmetry on cool
giant atmospheres has been investigated previously by \cite{scholz84} for giant
M and C stars, by \cite{plez92} for M giants, and by \cite{jjn92} for Carbon
stars. Our model grid extends this work toward warmer stars and reduces the gap
to the stars considered in \cite{epscma}. In addition, improved molecular line
lists are now available and since \cite{epscma} found a close connection
between line blanketing and sphericity for hot stars, the improved molecular
line data could affect the results.

In Fig.~\ref{change1_abs} we compare normalized
(to the same area) low-resolution ($20\ang$) spectra for 3 sets of models. Each
panel displays a NG-giant synthetic spectrum (full lines) and a synthetic
spectrum calculated with the same parameters and input physics, but assuming
plane parallel geometry.  In this figure, only the model with $\Teff=4000\K$
shows visible differences between the plane parallel and spherical
cases. However, plotting the relative differences of the spectra in
Fig.~\ref{change1} reveals systematic differences even for low resolution
spectra.  For $\Teff=3000\K$ (top panel), we find that the  spherical model
emits more flux than the plane parallel model in the short wavelength regime.
There are also some changes in the slopes of the spectra between $0.45$ and
$0.55\,\mu$m, which are hardly noticeable in the low resolution spectra. The
middle panel shows the results for $\Teff=4000\K$. This model displays a
different systematic change in the blue spectral region, the long wavelength
part of the displayed wavelength interval does not reveal large systematic
effects.  The highest effective temperature shown, $\Teff=5600\K$, shows the
smallest systematic changes. 

Figure \ref{change1a} is similar to the previous plot but it shows the changes
for a resolution of $2\ang$ (10 times better). From this figure it becomes
clear that the changes are due to the effect of line blanketing: line overlap
and individual line strengths change between spherical and plane parallel
models. In low resolution spectra this produces a change in the
pseudo-continuum that is formed by the overlapping spectral lines. In the
$\Teff=3000\K$ model the TiO lines are significantly affected.  This shows that
spectral analyses based on high-resolution spectra need to account for spherical
effects.

For longer wavelengths we find that the differences between spherical and plane
parallel models  are smaller. Figures \ref{change2} and
\ref{change2a} demonstrate this for low resolution and medium resolution
spectra, respectively. The changes are mostly due to the sensitivity of the CO
lines to changes in the structure of the atmosphere, therefore, they are larger
at lower effective temperatures. The figure with the higher resolution shows
that there are significant changes in individual water lines for low
$\Teff$. At higher effective temperatures, individual atomic lines are up to
15\% stronger in the plane parallel models (less emitted flux, since these are
absorption lines). The conclusion is that high-resolution spectral
analysis requires spherical models, whereas low resolution spectra or colors
can be interpreted using simple plane parallel models. 

In order to demonstrate the reason for these differences, we display in
Fig.~\ref{t-change} the electron temperatures as functions of our standard
optical depth $\tstd$, which is simply the optical depth in the continuum (b-f
and f-f processes) at a reference wavelength of $1.2\,\mu$m. The temperature
structures for the 2 models with $\Teff=5600\K$ are nearly identical, the
differences are less than about $80\K$ everywhere. In the model with the lowest
effective temperature ($3000\K$) the differences are larger, close to $200\K$
at $\tstd \approx 10^{-4}$. This amounts to nearly 10\% change in the absolute
temperature, which is quite significant for the formation of the spectrum. Note
that the line optical depth in the blue spectral region can be orders of
magnitude larger than $\tstd$ and many ``blue lines'' actually form around
$\tstd\approx 10^{-4}$.  The model with $\Teff=4000\K$ lies in between the two
extremes and is not shown in the figure. The temperature structure is only a
part of the full structure of the atmosphere, gas pressures and partial
pressures change accordingly.  For larger gravities, the differences between
plane parallel and spherical models become smaller and for gravities
$log(g)>3.0$ the differences are small enough to be negligible for most
applications.

\subsection{Spectra}

The changes of the synthetic spectra with gravity are illustrated in
Fig.~\ref{logg-change}. The differences are larger for the cooler model in the
sense that at lower $\Teff$, higher gravity model atmospheres emit more optical
flux than lower gravity models. This is caused by the enhanced near-IR
absorption of water vapor in the models with larger gravity and hence larger gas
pressures in the line forming region.  For larger effective temperatures
the differences diminish and are confined to individual gravity sensitive
features in the spectrum.

Changes in the metallicity $\mh$ result in changes of the synthetic spectra
that depend strongly on the effective temperature, as shown in Fig.~\ref{z-change}.
For high effective temperatures, the changes are small in the $\mh$ range we
have investigated.  At lower $\Teff$, however, the changes are dramatic as
illustrated in the top panel of Fig.~\ref{z-change}. This is due to the
increasing importance of molecules at lower temperatures. The concentration of
molecules and thus their opacities depend strongly on the metal abundances.
Molecules are less important at higher $\Teff$ and thus the spectra are less
sensitive to metallicity changes. 

The very small sensitivity of the atmospheric structure on the mass of the star
is mirrored by only small  changes in the resulting low-resolution spectra
(Fig.~\ref{sp-mass}). The differences correspond
to scaling factors close to unity over a large wavelength range
and are thus basically negligible for most purposes. The absolute changes
in limited wavelength intervals of the spectra, however, can be equivalent
to changes in $\logg$, thus the mass of the star needs to be considered as 
a parameters for applications that use absolute spectra.

\subsection{NLTE effects}

We have calculated a small number of NLTE models in order to investigate the
importance of NLTE effects on the structure
of the model atmospheres. The results for  cooler models were discussed
in \cite{tipap} and are not repeated here. Figures \ref{nlte40} and \ref{nlte56} show an overview
of selected NLTE species for models with $\Teff=4000\K$ and $5600\K$ for
$\logg=0.0$ and solar abundances. The total number of NLTE levels in
each model is 4532 with a total of 47993 primary NLTE lines 
\cite[see][and references therein for details]{jcam}. The following species
(and number of levels) were treated in NLTE:
H~I                         (30),
Mg~I                       (273),
Mg~II                       (72),
Ca~I                       (194),
Ca~II                       (87),
Fe~I                       (494),
Fe~II                      (617),
O~I                         (36),
O~II                       (171),
Ti~I                       (395),
Ti~II                      (204),
C~I                        (228),
C~II                        (85),
N~I                        (252),
N~II                       (152),
Si~I                       (329),
Si~II                       (93),
S~I                        (146),
S~II                        (84),
Al~I                       (111),
Al~II                      (188),
K~I                         (73),
K~II                        (22),
Na~I                        (53), and
Na~II                       (35).
For most of the species, the departure coefficients are always close
to unity, in particular for species with resonance lines and 
photoionization edges in the UV part of the spectrum. The species
shown in Figure \ref{nlte40} and \ref{nlte56} are the ones with the 
most pronounced departures from LTE. The departures are
generally too small to significantly affect the structure of the atmospheres.
However, NLTE effects do change the profiles of individual lines as shown
in \cite{tipap} for Ti~I lines. Therefore, abundance analyses of 
individual elements should take NLTE effects into account whenever 
possible.

\section{Summary and Conclusions}

In this paper we presented the extension of our grid of dwarf model atmospheres
\cite[]{ng-hot} to giant stars. The models were computed using spherical
symmetry, including spherical radiative transfer. We discuss the results of the
calculations. We have calculated a small number of NLTE models 
and verified the conclusions of \cite{tipap}: In the considered range
of effective temperatures, NLTE effects do not alter the 
structure of the atmospheres significantly but they can change the profiles
of individual lines. NLTE should therefore be considered in abundances
analyses. Our grid has been used successfully in evolutionary models of
Cepheids \cite[]{alibert99}. We provide the model structures and spectra
through the WWW and anonymous FTP for general use, see
{\tt http://dilbert.physast.uga.edu/\tilt yeti} or {\tt ftp://calvin.physast.uga.edu/pub/NG-giant}. 
In the next version of the grid we will use improved molecular line
opacities as well as dust opacities for low effective temperatures. In
addition, stellar winds models are needed for those models that are unstable
against radiation pressure, which occurs at both high and low effective
temperatures. 

\acknowledgments
This work was supported in
part by the CNRS and by NSF grant AST-9720704, NASA ATP grant NAG 5-3018 and LTSA grant NAG
5-3619 to the University of Georgia, and NASA LTSA grant NAG5-3435
and NASA EPSCoR grant NCCS-168  to Wichita
State University. 
This work was also supported in part by NSF grants AST-9417242, AST-9731450,
and NASA grant NAG5-3505; an IBM SUR grant to the University of Oklahoma.
Some of the calculations presented in this paper were
performed on the IBM SP2 and the SGI Origin 2000 of the UGA UCNS and on the IBM
SP of the San Diego Supercomputer Center (SDSC), with support from the National
Science Foundation, and on the Cray T3E of the NERSC with support from the DoE.
We thank all these institutions for a generous allocation of computer time.

\clearpage

\bibliography{yeti,opacity,mdwarf,radtran,general,opacity-fa,mdwarf-fa}

\begin{thebibliography}{}

\bibitem[Alibert { et~al.}(1999)Alibert, Baraffe, Hauschildt, and
  Allard]{alibert99}
Alibert, Y., Baraffe, I., Hauschildt, P.~H., \& Allard, F. 1999,
\newblock A\&A, in press.

\bibitem[Aufdenberg { et~al.}(1998a)Aufdenberg, Hauschildt, Shore, and
  Baron]{epscma}
Aufdenberg, J.~P., Hauschildt, P.~H., Shore, S.~N., \& Baron, E. 1998a,
\newblock ApJ, 498, 837.

\bibitem[Aufdenberg { et~al.}(1998b)Aufdenberg, Hauschildt, Sankrit, and
  Baron]{betacma}
Aufdenberg, J.~P., Hauschildt, P.~H., Sankrit, R., \& Baron, E. 1998b,
\newblock MNRAS, in press.

\bibitem[Baron \& Hauschildt(1998)Baron and Hauschildt]{parapap2}
Baron, E., \& Hauschildt, P.~H. 1998,
\newblock ApJ, 495, 370.

\bibitem[Baschek, Scholz, \& Wehrse(1991)Baschek, Scholz, and Wehrse]{bsw91}
Baschek, B., Scholz, M., \& Wehrse, R. 1991,
\newblock A\&A, 246, 374.

\bibitem[Hauschildt, Allard, \& Baron(1999)Hauschildt, Allard, and
  Baron]{ng-hot}
Hauschildt, P., Allard, F., \& Baron, E. 1999,
\newblock ApJ, 512, 377.

\bibitem[Hauschildt \& Baron(1999)Hauschildt and Baron]{jcam}
Hauschildt, P.~H., \& Baron, E. 1999,
\newblock Journal of Computational and Applied Mathematics, in press.

\bibitem[Hauschildt { et~al.}(1997)Hauschildt, Allard, Alexander, and
  Baron]{tipap}
Hauschildt, P.~H., Allard, F., Alexander, D.~R., \& Baron, E. 1997,
\newblock ApJ, 488, 428.

\bibitem[Hauschildt, Baron, \& Allard(1997)Hauschildt, Baron, and
  Allard]{parapap}
Hauschildt, P.~H., Baron, E., \& Allard, F. 1997,
\newblock ApJ, 483, 390.

\bibitem[Jaschek \& Jaschek(1995)Jaschek and Jaschek]{jaschek95}
Jaschek, C., \& Jaschek, M. 1995,
\newblock {\em The behavior of chemical elements in stars\/},
\newblock Cambridge University Press.

\bibitem[J{\o}rgensen, Johnson, \& Nordlund(1992)J{\o}rgensen, Johnson, and
  Nordlund]{jjn92}
J{\o}rgensen, U.~G., Johnson, H.~R., \& Nordlund, A. 1992,
\newblock A\&A, 261, 263.

\bibitem[Mihalas(1970)Mihalas]{mihalas70}
Mihalas, D. 1970,
\newblock {\em Stellar Atmospheres\/},
\newblock Freeman, San Francisco, 1 edition.

\bibitem[Plez, Brett, \& Nordlund(1992)Plez, Brett, and Nordlund]{plez92}
Plez, B., Brett, J.~M., \& Nordlund, A. 1992,
\newblock A\&A, 256, 551.

\bibitem[Scholz \& Tsuji(1984)Scholz and Tsuji]{scholz84}
Scholz, M., \& Tsuji, T. 1984,
\newblock A\&A, 130, 11.

\bibitem[Uns{\"o}ld(1968)Uns{\"o}ld]{unsoeld55}
Uns{\"o}ld, A. 1968,
\newblock {\em Physik der Sternatmosph{\"a}ren\/},
\newblock Springer Verlag, Heidelberg, 2 edition.

\end{thebibliography}

\clearpage
\section{Figures}

\begin{figure}[b]
\caption[]{\label{T-struc-logg} Temperature structures for solar abundance models with 
$\Teff=5600\K$, $4200\K$, and $3000\K$  for varying gravities (as indicated).
$\tstd$ is the optical
depth in the continuum (b-f and f-f processes) at a wavelength of $1.2\,\mu$m. 
}
\end{figure}

\begin{figure}[t]
\caption[]{\label{extension} Radial extensions $R_{\rm out}/R_{\rm in}$ 
for solar abundance models with 
$\Teff=5600\K$, $4200\K$, and $3000\K$  (as indicated).
}
\end{figure}

\begin{figure}[t]
\caption[]{\label{change1_abs} Comparison between solar abundance giant models
calculated using spherical geometry and radiative transfer (full curves) and
plane parallel  geometry and radiative transfer(dotted curves) in the
blue/optical spectral region. The resolution of the spectra has been reduced to 
$20\ang$.}
\end{figure}

\begin{figure}[t]
\caption[]{\label{change1} Relative flux change between spherical and plane parallel
model calculations. The y-axis shows $(f_p-f_s)/f_s$, where $f_p$ is the flux of 
the plane parallel model and $f_s$ is the flux calculated for the spherical model.
The resolution of the spectra was reduced to $20\ang$.
}
\end{figure}

\begin{figure}[t]
\caption[]{\label{change1a} Relative flux change between spherical and plane parallel
model calculations. The y-axis shows $(f_p-f_s)/f_s$, where $f_p$ is the flux of 
the plane parallel model and $f_s$ is the flux calculated for the spherical model.
The resolution of the spectra is $2\ang$.
}
\end{figure}

\begin{figure}[t]
\caption[]{\label{change2} Relative flux change between spherical and plane parallel
model calculations. The y-axis shows $(f_p-f_s)/f_s$, where $f_p$ is the flux of 
the plane parallel model and $f_s$ is the flux calculated for the spherical model.
The resolution of the spectra was reduced to $20\ang$.
}
\end{figure}

\begin{figure}[t]
\caption[]{\label{change2a} Relative flux change between spherical and plane parallel
model calculations. The y-axis shows $(f_p-f_s)/f_s$, where $f_p$ is the flux of 
the plane parallel model and $f_s$ is the flux calculated for the spherical model.
The resolution of the spectra is $2\ang$.
}
\end{figure}

\begin{figure}[t]
\caption[]{\label{t-change} Temperature structures for spherical (full curves)
and plane parallel (dotted lines) model calculations. $\tstd$ is the optical
depth in the continuum (b-f and f-f processes) at a wavelength of $1.2\,\mu$m. 
}
\end{figure}

\begin{figure}[t]
\caption[]{\label{logg-change} Sensitivity of the synthetic spectra to gravity
changes for solar abundance models with $\Teff=5600\K$, $4200\K$, and $3000\K$
(as indicated). The resolution of the spectra has been reduced to $20\ang$.
}
\end{figure}

\begin{figure}[t]
\caption[]{\label{z-change} Sensitivity of the synthetic spectra to metallicity
changes for models with $\Teff=5600\K$, $4200\K$, and $3200\K$ 
(as indicated) and $\logg=0.0$. For the $\Teff=3200\K$ model the metallicities 
$\mh=0.0$ and $\mh=-0.3$ are shown whereas for the hotter two models the 
metallicities shown are $\mh=0.0$ and $\mh=-0.7$.
The resolution of the spectra has been reduced to $20\ang$.
}
\end{figure}

\begin{figure}[t]
\caption[]{\label{sp-mass} Synthetic spectra at $20\ang$ resolution
for models with $\Teff=3600\K$, $\logg=0.0$, solar abundances, and 
stellar masses of $7.5\Msun$ and $2.5\Msun$.
}
\end{figure}

\begin{figure}[t]
\caption[]{\label{nlte40} Overview over selected departure coefficients
for a NLTE model with $\Teff=4000\K$, $\logg=0.0$, and solar abundances.
}
\end{figure}

\begin{figure}[t]
\caption[]{\label{nlte56} Overview over selected departure coefficients
for a NLTE model with $\Teff=5600\K$, $\logg=0.0$, and solar abundances.
}
\end{figure}

\end{document}